\documentclass[12pt,a4paper]{article}
\usepackage[pdftex]{graphicx}
\usepackage{color}
\usepackage[indentfirst,compact,topmarks,calcwidth,pagestyles]{titlesec}
\usepackage{titletoc}
\usepackage{sectsty}
\usepackage{amsmath} 
\usepackage{cancel}
\usepackage[noadjust]{cite}
\usepackage{lineno}
\usepackage[top=10mm,bottom=20mm,left=30mm,right=20mm,includehead,head=12pt,headsep=0cm,nofoot]{geometry}
\sectionfont{\fontsize{12}{14}\selectfont}
\subsectionfont{\fontsize{12}{14}\selectfont}

\numberwithin{equation}{section} 

\usepackage[T2A]{fontenc}
\usepackage[utf8]{inputenc}
\begin{document}
draft 0.1
\vskip 15mm
\begin{center}
{\bfseries Approach to relativistic quark confinement potentials from QCD}
\vskip 5mm
IvanYeletskikh$^{a,*}$\vskip 5mm
\end{center}
$^{a}$ Joint Institute for Nuclear Research, Dubna, Russia; \\
$^{*}$ email: ivaneleckih@jinr.ru
\vskip 10mm
\centerline{\bf Abstract}
We present an attempt towards constructing relativistic quark-gluon quasi-potentials from QCD using Unitary Clothing Transformations (UCT) approach. The brief review of the UCT method is given. Starting from a simplified model of QCD hamiltonian ($H$), 
we define the 'color-clothed' representation (CCR) of $H$. In this representaion, it is required that only color-singlet states are the eigenstates of the $H$.
In CCR, $H$ contains the family of quantum relativistic potentials acting between 'color clothed' quarks and gluons. Simultaneously, newly defined quarks acquire new masses (expected to be close to constituent masses), and new renormalized couplings. Properties of 'color clothed' quark-gluon potentials are consistent with the concept of 'color confinement' in a natural way.

The presented paper is a draft, aiming in the first instance for discussion on the approach presented and for possible criticism as well.
In the current version we  discuss the basics of the approach and, as an example, derive explicitely one of the operators, responsible for the lowest order quark-quark relativistic confinement potential.
\vskip 10mm

\section{\label{sec:introduction}Introduction}
The conventional Quantum Field Theory (QFT) proved itself as an effective theoretical tool. However, despite the great accuracy this theory demonstrates in describing many of the observable effects, early at the stage of it's creation there was realized the presence of some serious conceptual problems. Created as a theory, operating in the infinite-dimensional space of particle states, the QFT revealed some badly undesirable properties, in particular, the persistent effects of quanta self-interactions, producing infinities in observable results.
Overcoming these problems required creating tremendous mathematical apparatus called renormalization. Assigning new masses and coupling to bare quanta allowed to obtain the finite S-matrix describing quantum relativistic phenomena, including creation and annihilation of particles.
However, soon became recognition that such procedure is nothing but moving the problem from one place to another. It was stressed, 
that conventional renormalization procedure had some fundamental defects and, in fact, left the theory quanta essentially ''bare'', in particular, the dynamics of these particles, e.g., time evolution or inertial transformations, was described by infinite unobservable Hamiltonian and boost operators.

The concept of the Clothed Particle Representation (CPR) was introduced by Greenberg and Schweber \cite{greenberg_1} in 1950-th, as an attempt towards solving the above mentioned difficulties of QFT. They suggested redefinition of the creation/annihilation operators of particles ($\alpha$) via unitary transformation $\alpha_{c}=U^{-1}\alpha U$, demanding the new vacuum and one-particle states (contrary to the ''bare'' ones) to be the eigenstates of the total Hamiltonian operator of the theory:
\begin{equation}\label{CPR1}
\begin{split}
H\left| 0_{c}\right\rangle &= 0, \\ H\left| \alpha_{c}\right\rangle &= E\left| \alpha_{c}\right\rangle,
\end{split}
\end{equation}
where $\left| \alpha_{c}\right\rangle$ -- so-called ''clothed'' one-particle states, $\left| 0_{c}\right\rangle$ - ''clothed'' vacuum state, $H$ -- total (free part plus interaction) Hamiltonian. 

In CPR, the Hamiltonian (and other Poincare generators) is defined in terms of relativistic interactions bettween 'clothed' particles, is free of divergences and produces finite 'physical' S-operator.
It is important to be aware of the fact, that transformation $U$, acting on particle operators, formally leaves Hamiltonian operator and thus the S-matrix intact. All divergences, associated with 'bare' particles' effects, are 'accumulated' into new creation/annihilation operators of 'clothed' particles.

Saying the same in more formal way, the original field-theoretical interaction (for most of realistic theories), acting on ''bare'' particles, violates (\ref{CPR1}), and thus produce undesirable observable properties of 'bare' particles. Being expressed in terms of 'clothed' creation/annihilation operators, Hamiltonian looks as a set of interaction operators between clothed particles, that satisfy (\ref{CPR1}).

Further development of the UCT method, related approaches and their applications to the different theories, such as QED, mesodynamics or even some toy gravity models, has been implemented in works of Shirikov, Shebeko, Stefanovich, etc. (see, e.g., \cite{shebeko_0, shebeko_1, shebeko_2, stefanovich_1}).
We suggest reader to get acquainted with these works, and in what follows, we will avoid discussing the concepts of UCT method in detail. Instead, we will focus on the idea of it's application to the QCD.

\section{\label{sec:UCT0}Application of UCT method to QCD}
In case of realistic theories of electromagnetic, weak interactions or some similar toy models, the physical (observable) states are usually associated with one-particle (electron, photon, etc.) states. 
In the framework of UCT method, the natural idea was to require the one-particle states to be the eigenstates of the total Hamiltonian operator.
In case of QCD, the single quarks and gluons are unobservable, being confined via strong color interactions inside color-singlet (we also will call them 'color-neutral') bound states.
For an effective theory one should consider only 'color-neutral' states as observable, and thus -- in the context of UCT approach -- to require them to be eigenstates of QCD hamiltonian.

In case of QCD, according to the line of argument presented above, we call 'bad' those operators that have empty energy shell \textit{or} translate 'colored' states to 'colored' states. 
Besides operators carrying transitions from vaccum and one-particle states to states with two or more particles, this definition assigns label 'bad' to some of the operators with 'good' (physical) particle kinematics, i.e., operators with non-empty energy shell. The examples from the latter class of operators are $qq\to qq$, $gq\to qq\bar{q}$, $q\bar{q}\to q\bar{q}$ (in the last case, $q$ and $\bar{q}$ in the initial and final states must possess different color/anticolor).
Quarks and gluons that are 'grouped' within the color-neutral states and participate only in interactions between 'color-neutral' states, will be called 'color clothed' quarks and gluons. The representation of physical operators in terms of 'color clothed' quark/gluons we will refer to as 'color clothed' representation.

\subsection{\label{sec:H0}QCD Hamiltonian}
As a starting point we will consider the field theoretical QCD Hamiltonian (e.g., ref. \cite{brodsky_1}) in the instant form of relativistic dynamics:
\begin{equation}\label{field_hamiltonian0}
\begin{split}
H = \int{dx(F^{0\mu}_{a}{{F_{\mu}}^{0}}^{a} + \frac{1}{4}F^{\mu\nu}_{a}F_{\mu\nu}^{a} - \frac{1}{2}[\bar{\psi_{c}}(i\gamma^{\mu}D^{cc'}_{\mu}-m)\psi + h.c.] + \frac{1}{2}[i\bar{\psi_{c}}\gamma^{0}D^{cc'}_{0}\psi_{c'} + h.c.])},
\end{split}
\end{equation}
where:
\begin{itemize}
\item $F_{\mu\nu}^{a} \equiv \partial_{\mu}A_{\nu}^{a} - \partial_{\nu}A_{\mu}^{a} + gf^{abs}A_{\mu}^{b}A_{\nu}^{s}$ -- gluon field tensor. Indices $a, b, s$ vary from 1 to 8 and count gluon fields, $f^{abs}$ -- antisymmetric structure functions, explicit form of which won't be important for our study;
\item $A_{\mu} = T_{cc'}^{a}A_{\mu}^{a}$. Indices $c, c' = 1,2,3$ count quark colors;
\item $D_{\mu}^{cc'}\equiv\delta^{cc'}\partial^{\mu} + igT_{a}^{cc'}A_{\mu}^{a}$ -- covariant derivative;
\item $T_{cc'}^{a}$ are hermitian and traceless 3$\times$3 matrices, the explicit forms for them can be seen from (\ref{gluon_field0});
\item all field operators are locally defined in Minkowsky spacetime, so everywhere in (\ref{field_hamiltonian0}) we assume $\psi \equiv \psi(x), A^{\mu} \equiv A^{\mu}(x)$;
\item we consider only one quark flavor for simplicity;
\end{itemize}
\begin{equation}\label{gluon_field0}
\begin{split}
T^{cc'}_{a}A^{a}_{\mu} = \frac{1}{2}\begin{pmatrix}
\frac{1}{\sqrt{3}}A_{\mu}^{8} + A_{\mu}^{3} & A_{\mu}^{1} - iA_{\mu}^{2}  & A_{\mu}^{4} - iA_{\mu}^{5} \\
A_{\mu}^{1} + iA_{\mu}^{2} & \frac{1}{\sqrt{3}}A_{\mu}^{8} - A_{\mu}^{3} & A_{\mu}^{6} - iA_{\mu}^{7} \\         
A_{\mu}^{4} + iA_{\mu}^{5} & A_{\mu}^{6} - iA_{\mu}^{7} & -\frac{2}{\sqrt{3}}A_{\mu}^{8}
\end{pmatrix},
\end{split}
\end{equation}
Hamiltonian density (\ref{field_hamiltonian0}) contains the part describing free quarks and gluons, the vertex 3-linear quark-gluon interaction, 3- and 4-linear gluon-gluon interactions  (ref. \cite{brodsky_1}) and also Coulomb-like potentials for quarks and gluons (see \cite{weinberg_1}).
For now, we only will take into account the 3-linear quark-gluon coupling:
\begin{equation}\label{field_hamiltonian1}
\begin{split}
V = \int{gJ^{\mu}_{a}(x)A_{\mu}^{a}(x)dx},
\end{split}
\end{equation}
where $J^{\mu}_{a}$ is the quark color curent $J^{\mu}_{a}=\bar{\psi_{c}}\gamma^{\mu}T^{cc'}_{a}\psi_{c'}$.

For later usage, we also present the interaction operator (\ref{field_hamiltonian1}) in terms of particle creation/annihilation operators:
\begin{equation}\label{field_hamiltonian11}
\begin{split}
V =& \frac{g}{(2\pi)^{3/2}}\int{\frac{d^{3}pd^{3}qd^{3}k}{(E_{\vec{p}}E_{\vec{q}}\omega_{\vec{k}})^{1/2}}\bar{U}_{i}(\vec{p},s) T^{a}_{cc'}\gamma_{\mu}{U}_{j}(\vec{q},r)} \times \\
& \times{F^{c}_{i}}^{\dagger}(\vec{p},s) {F^{c'}_{j}}(\vec{q},r) {\epsilon^{\mu}}^{*}(\vec{k},\lambda)a^{\dagger}_{a}(\vec{k},\lambda) \delta(\vec{p}-\vec{q}+\vec{k}) + h.c.,
\end{split}
\end{equation}
where:
\begin{itemize}
\item $F_{c}(\vec{p},s) = \begin{pmatrix} b_{c}(\vec{p},s) \\ d_{c}^{\dagger}(-\vec{p},s) \end{pmatrix}$, $F_{c}^{\dagger}(\vec{p},s) = ( b_{c}^{\dagger}(\vec{p},s) , d_{c}(-\vec{p},s))$, $b_{c}^{\dagger}(\vec{p},s) (b_{c}(\vec{p},s))$ and $d_{c}^{\dagger}(\vec{p},s) (d_{c}(\vec{p},s))$ are creation (annihilation) operators of quarks and antiquarks with momentum $\vec{p}$, spin projection $s$ and color index $c$, respectively;
\item $a^{\dagger}_{a}(\vec{k},\lambda) (a_{a}(\vec{k},\lambda))$ --  creation (annihilation) operator of gluon with momentum $\vec{k}$, spin projection $\lambda$ and gluon index $a$;
\item commutation relations of quark and gluon operators are standard for bosonic and fermionic operators: $[a_{a_1}(\vec{k}_{1},\lambda_{1}),a^{\dagger}_{a_2}(\vec{k}_{2},\lambda_{1})]$ $ = \delta_{a_{1}a_{2}}\delta_{\lambda_{1}\lambda_{2}}\delta(\vec{k_1}-\vec{k_1})$, \\ $\{F^{c_{1}}_{i_{1}}(\vec{p}_{1},s_{1}),{F^{c_{2}}_{i_{2}}}^{\dagger}(\vec{p}_{2},s_{2})\}$ $ = \delta_{c_{1}c_{2}}\delta_{s_{1}s_{2}}\delta_{i_{1}i_{2}}\delta(\vec{p_1}-\vec{p_1})$;
\item $\bar{U}(\vec{p},s) = \begin{pmatrix} \bar{u}(\vec{p},s) \\ v(-\vec{p},s) \end{pmatrix}$, $U(\vec{p},s) = ( u(\vec{p},s), \bar{v}(-\vec{p},s) )$ are Dirac spinors;
\item $\epsilon^{\mu}(\vec{k},\lambda)$ is gluon coefficient function;
\item $E_{\vec{p}} = \sqrt{m^{2}+\vec{p}^{2}}$, $\omega_{\vec{k}} = |\vec{k}|$ -- energy of quark, gluon;
\end{itemize}

\subsection{\label{sec:UCT1}Removing 'bad' terms}
As stressed before -- representing field-theoretical QCD Hamiltonian in terms of 'color clothed' quark/gluon creation/annihilation operators consists in removing 'bad' operators from it order-after-order in coupling. The quark-gluon interaction (\ref{field_hamiltonian1})  is 'bad' according to our definition, since it has empty energy shell. In the most general form, the unitary transformation of creation/annihilation operators looks as follows:
\begin{equation}\label{uct0}
\begin{split}
\alpha_{c}=W^{-1}\alpha W,
\end{split}
\end{equation}
where $\alpha$ is the set of initial particle creation/annihilation operators, and $\alpha_c$ is the set of new, in our case 'color clothed' particle creation/annihilation operators, $W$ is the operator of unitary transformation.
Operator defined in terms of operators $\alpha$ could be expressed via set of operators $\alpha_{c}$ in the following manner:
\begin{equation}\label{uct1}
\begin{split}
O(\alpha) = W(\alpha_{c})O(\alpha_{c})W^{\dagger}(\alpha_{c}) = e^{R(\alpha_{c})}O(\alpha_{c})e^{-R(\alpha_{c})},
\end{split}
\end{equation}
where $R$ is the generator of unitary transformation $W=e^{R}$.
The expression (\ref{uct1}) could be unfolded into a set of multiple commutators, so that we obtain:
\begin{equation}\label{uct1}
\begin{split}
O(\alpha) = O(\alpha_c) + [R(\alpha_{c}),O(\alpha_{c})] + \frac{1}{2!}[R(\alpha_{c}), [R(\alpha_{c}),O(\alpha_{c})]] + ...,
\end{split}
\end{equation}
Thus, for the Hamiltonian operator, it's expression in terms of 'clothed' particle operators looks as follows:
\begin{equation}\label{uct11}
\begin{split}
H(\alpha) = H(\alpha_c) + [R(\alpha_{c}),H(\alpha_{c})] + \frac{1}{2!}[R(\alpha_{c}), [R(\alpha_{c}),H(\alpha_{c})]] + ...,
\end{split}
\end{equation}
Splitting Hamiltonian into free part $H_{0}$ and interaction $V$, we get from (\ref{uct11}):
\begin{equation}\label{uct2}
\begin{split}
H(\alpha) = & H_0(\alpha_c) + V(\alpha_c) + [R(\alpha_{c}),H_0(\alpha_{c})] + \\ &
[R(\alpha_{c}),V(\alpha_{c})] + \frac{1}{2!}[R(\alpha_{c}), [R(\alpha_{c}),H_0(\alpha_{c})]] + \frac{1}{2!}[R(\alpha_{c}), [R(\alpha_{c}),V(\alpha_{c})]] +...,
\end{split}
\end{equation}

\begin{figure}[!htb]\center
\includegraphics[width=0.28\textwidth]{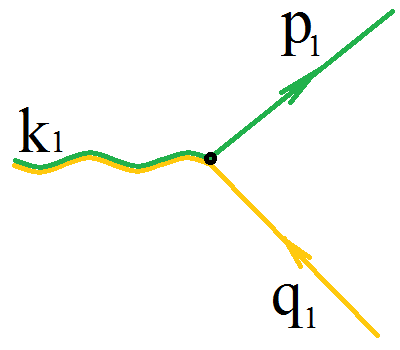}      \hskip 20mm
\includegraphics[width=0.28\textwidth]{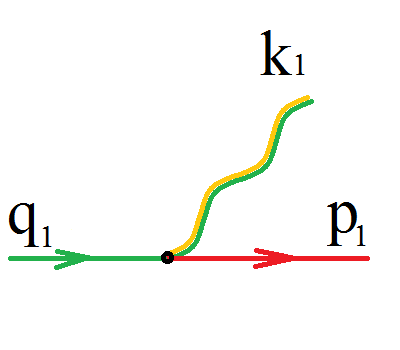}
\caption{ \label{bad0}
Examples of 'bad' terms in Hamiltonian (\ref{uct2}) in the 1st order in coupling. Quark operators are displayed as straight lines, gluons -- as wavy lines. Colors illustrate color charges of quarks, antiquarks and gluons (yellow color displays anti-red color). Direction of quark line distinguishes quarks (left-to-right arrows) from anti-quarks (right-to-left arrows). Both operators have empty energy shell \textit{and} describe transitions between 'colored' states.}
\end{figure}

\begin{figure}[!htb]\center
\includegraphics[width=0.21\textwidth]{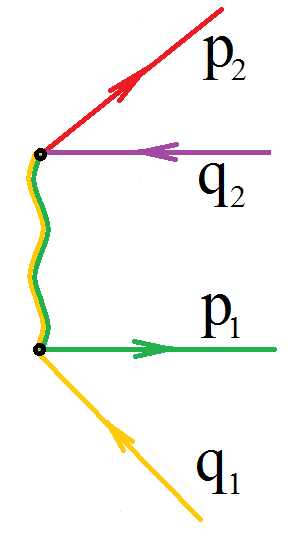} \hskip 20mm
\includegraphics[width=0.35\textwidth]{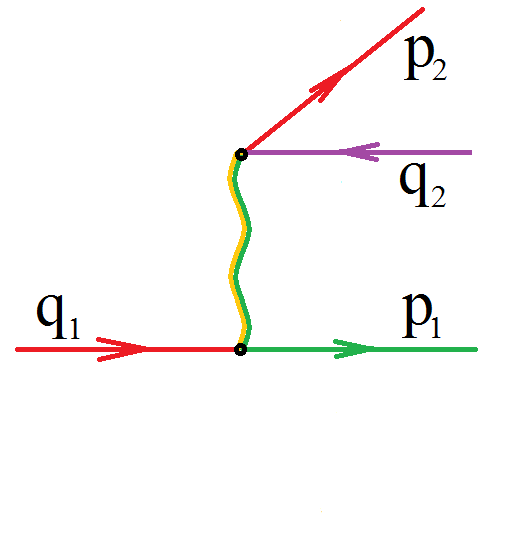}
\\
\vskip 5mm
\includegraphics[width=0.4\textwidth]{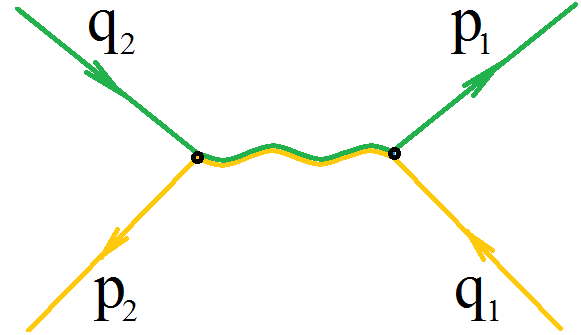}
\includegraphics[width=0.45\textwidth]{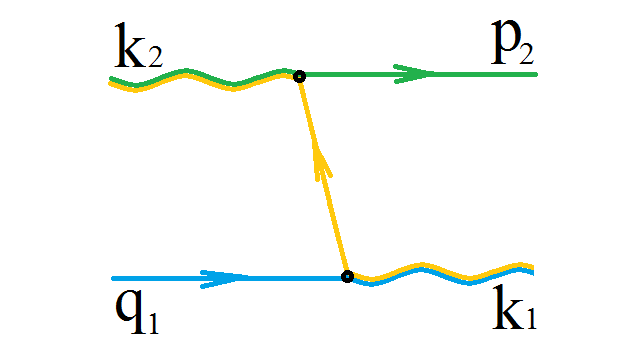}
\caption{ \label{bad1}
Examples of 'bad' terms in Hamiltonian (\ref{uct2}) in 2nd order in coupling. Yellow displays anti-red, purple displays anti-green. Top left diagram is an example of operator with empty energy shell, translating vacuum state to multi-quark color-neutral state. Top right diagram corresponds to operator with empty energy shell \textit{and} describing transitions between 'colored' stated. Bottom diagrams correspond to 'bad' operators with non-empty energy shell, but describing transitions between 'colored' states.}
\end{figure}

The clothed particle represetation (as defined in original work \cite{greenberg_1}) is achieved via defining $R$ in such way, that $H$ (expressed in terms of $\alpha_{c}$) does not contain any 'bad' terms.
In the example of QED Hamiltonian (or in similar realistic or toy theories), the 'bad' operators are defined as the operators with empty energy shell, i.e., have empty phase space where energy-momentum is conserved. These operators implement transition from either vacuum or one-particle states to multi-particle states and thus, prevent one-bare-particle states to be eigenstates of $H$.
In case of QCD Hamiltonian, we aim at 'color clothed' representation, thus, our definition of $R$ will require that $H$ does not contain any operators that have empty energy shell or translate 'colored' states into 'colored' states.
Examples of 'bad' operators in QCD Hamiltonian (\ref{uct2}) in 1st and 2nd orders in coupling (\ref{field_hamiltonian1}) are shown in Figures \ref{bad0}, \ref{bad1}.

Mathematically, procedure of defining $R$ is carried order by order in coupling. Having defined $R$ in the 1st order in coupling in such way, that $H$ do not contain 'bad' terms in 1st order -- we determine $H$ in 2nd order. This allows defining $R$ in 2nd order and $H$ in 3rd order, etc.
After defining $R$ -- the Hamiltonian operator will contain only 'good' interaction operators between newly defined quarks and gluons. One should here accent again, that Hamiltonian operator is formally not altered after transforming the particle operators and so, stays the same also the S-matrix.
\begin{figure}[!htb]\center
\includegraphics[width=0.45\textwidth]{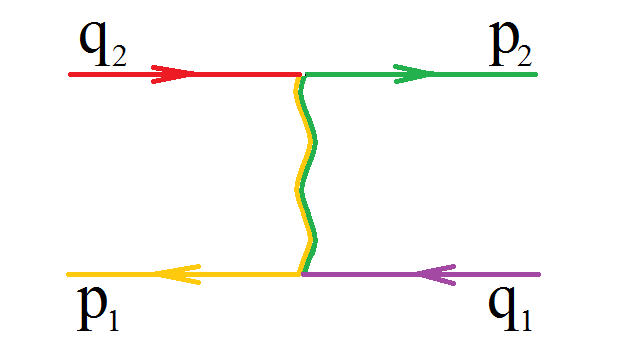} 
\includegraphics[width=0.45\textwidth]{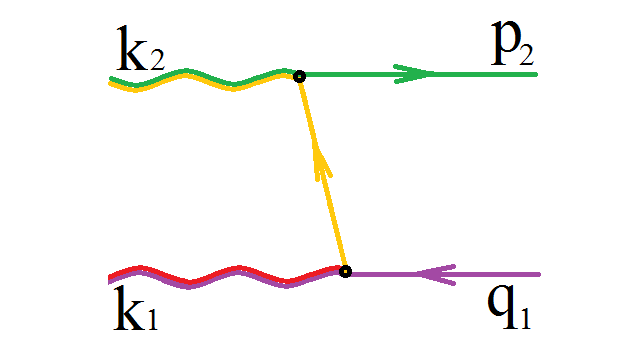}
\caption{ \label{good0}
Examples of 'good' terms in Hamiltonian (\ref{uct2}) in the 2nd order in coupling. As before, yellow displays anti-red, purple displays anti-green. Right diargam corresponds to process $q\bar{q}\to q\bar{q}$, left diagram -- to process $gg\to q\bar{q}$.}
\end{figure} 
\begin{figure}[!htb]\center
\includegraphics[width=0.55\textwidth]{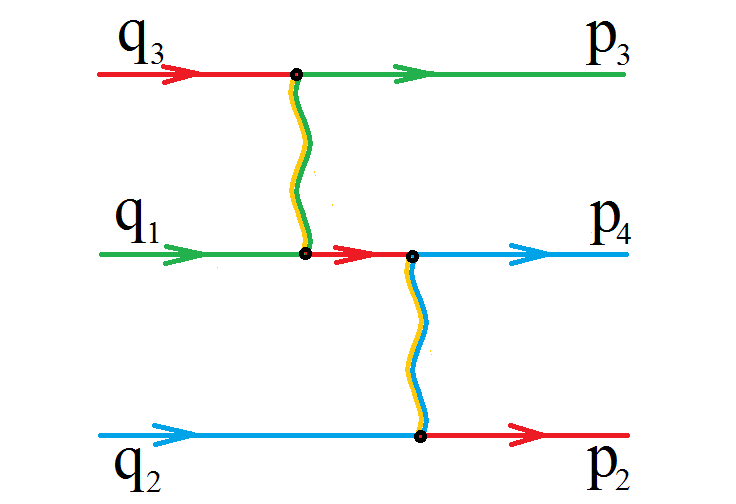} 
\caption{ \label{good1}
One of 'good' terms in Hamiltonian (\ref{uct2}) in the 4th order in coupling, corresponding to interaction in 3-quark sector. }
\end{figure} 

3-linear quark-gluon interaction (\ref{field_hamiltonian1}) totally consists of 'bad' terms, since all of 3-linear vertices have empty energy shell. That is why -- in our simplified QCD model -- we will determine $R$ in the first order in such way that $V$ is absent in $H$ (\ref{uct2}):

\begin{equation}\label{series0}
\begin{split}
V(\alpha_c) + [R^{(1)}(\alpha_{c}),H_0(\alpha_{c})] = 0,
\end{split}
\end{equation}
where $R^{(1)}$ is $R$ in the 1st order in coupling, $R=\sum_{k=1}^{\infty}R^{(k)}$.
After imposing requirement \ref{series0}, Hamiltonian (\ref{field_hamiltonian1}) acquires the following form:
\begin{equation}\label{uct3}
\begin{split}
H(\alpha) = & H_0(\alpha_c) + \frac{1}{2}[R^{(1)}(\alpha_{c}),V(\alpha_{c})] 
+ \frac{1}{3}[R^{(1)}(\alpha_{c}), [R^{(1)}(\alpha_{c}),V(\alpha_{c})]] + \\ & + \frac{1}{8}[R^{(1)}(\alpha_{c}), [R^{(1)}(\alpha_{c}), [R^{(1)}(\alpha_{c}),V(\alpha_{c})]]] +...,
\end{split}
\end{equation}
where only free part and terms depending on $R^{(1)}$ are shown.
Hamiltonian (\ref{uct3}) does not contain 'bad' terms in 1st order in coupling, however, it now does contain 'bad' terms in 2nd order, some of which are shown in Figure \ref{bad1}. All of them come from $\frac{1}{2}[R^{(1)},V]$ term.
In case we denote these terms as $\frac{1}{2}[R^{(1)}, V]_{bad}$ and 'good' terms as $\frac{1}{2}[R^{(1)}, V]_{good}$ -- the requirement on defining $R$ in the 2nd order in coupling is the following:
\begin{equation}\label{series1}
\begin{split}
[R^{(2)},H_0] + \frac{1}{2}[R^{(1)}(\alpha_{c}),V]_{bad} = 0.
\end{split}
\end{equation}
Here and after we omit showing dependence on $\alpha_c$ and assume all operators are defined in terms of 'color-clothed' quarks/gluons.
After having defined $R^{(2)}$ from (\ref{series1}), we obtain Hamiltonian in the form:
\begin{equation}\label{uct3}
\begin{split}
H = & H_0 + \frac{1}{2}[R^{(1)},V]_{good}+ \frac{1}{3}[R^{(1)}, [R^{(1)},V]] + \\ & + \frac{1}{8}[R^{(1)}, [R^{(1)}, [R^{(1)},V]]] + \frac{1}{2}[R^{(2)}, [R^{(1)},V]_{good}] + \frac{1}{4}[R^{(2)}, [R^{(1)},V]_{bad}] +...,
\end{split}
\end{equation}
where only free part and terms depending on $R^{(1)}$ and $R^{(2)}$ are shown.
Now we got rid of all 'bad' terms in the 2nd order in coupling. The next step would be to extract 'bad' terms in 3rd order and to define $R^{(3)}$...
However, we will not go that far, instead, we analyse Hammiltonian (\ref{uct3}), obtain examples of quark-gluon potentials in lower orders and foresee properties of the effective quark-gluon interactions in higher orders.

First of all, one stresses that 'good' terms in Hamiltonian (\ref{uct3}) are determined up to (and including) the 4th order in coupling, since in our model we do not have 'good' terms of the 1st order. 'Good' terms in 5th order depend already on $R^{(3)}$.
Thus, we will analyse (\ref{uct3}) only up to 4th order in coupling.

Second, since all 1st order terms are 'bad' in our model, just like in QED or mesodynamics models, the structure of $R^{(1)}$ generator is the same in our model and in QED/mesodynamics models. It means, that all quark-gluon interaction operators in Hamiltonian (\ref{uct3}) depending on $R^{(1)}$ and $V$ only have the same structure as in models described in \cite{shebeko_0, shebeko_1, shebeko_2, stefanovich_1}. This is completely true for 'color-clothed' quark-gluon interactions of the 2nd and 3rd orders in coupling. So -- for detailed discussions concerning the structure of Hamiltonian (\ref{uct3}) in the 2nd and 3rd orders in coupling -- we send reader back to works \cite{shebeko_0, shebeko_1, shebeko_2, stefanovich_1} and related links therein.

The differences between 'clothed particles' and 'color-neutral clothed states' representations emerges in the $R^{(2)}$ generator due to different definitions of 'bad' terms. Namely, contrary to 'clothed particles' approach, we have a set of 2nd order 'bad' quark-gluon interactions that have non-empty energy shell. We will see in the next part of this section, that generator $R^{(2)}$, eliminating such terms from $H$, will contain propagators, defining potentials, confining quarks/gluons within hadrons.

We expect, that higher order operators (starting from 4th) depending on $R^{(2)}, R^{(3)}, ...$, besides defining conventional relativistic scattering potentials, will contain additional structures, corresponding to relativistic confinement potantials and having relevant properties. It is also expected, that these new structures will interfere with scattering operators, leading to different from usual approaches results for QCD observables.

\subsection{\label{sec:UCT2}Quark-gluon confinement potentials from UCT method.}
As an example, in this subsection we will consider $3q\to3q$ potential in the 4th order in coupling (Fig. \ref{good1}). The operators, responsible for this process, come from the following terms: $\frac{1}{8}[R^{(1)}, [R^{(1)}, [R^{(1)},V]]]$, $\frac{1}{4}[R^{(2)}, [R^{(1)},V]_{bad}]$ in Hamiltonian (\ref{uct3}). 
The term $\frac{1}{8}[R^{(1)}, [R^{(1)}, [R^{(1)},V]]]$ in it's 'good' part contains scattering potentials, consistent with the results of conventional QFT.
Besides this, the terms $\frac{1}{4}[R^{(2)}, [R^{(1)},V]_{bad}]$ and $\frac{1}{2}[R^{(2)}, [R^{(1)},V]_{good}]$ also contain contributions to scattering potentials.
However, in the first instance, we are interested in the term $\frac{1}{4}[R^{(2)}, [R^{(1)},V]_{bad}]$, since this term in our scheme is expected to generate some of the lowest order confinement potentials.

For the purpose of deriving quark-gluon potentials, we need to define explicitely for our model generators $R^{(1)}$, $R^{(2)}$.
From equation (\ref{series0}), the $R^{(1)}$ is found in the form:
\begin{equation}\label{field_hamiltonian3}
\begin{split}
R^{(1)} =& \frac{g}{(2\pi)^{3/2}}\int{\frac{d^{3}pd^{3}qd^{3}k}{(E_{\vec{p}}E_{\vec{q}}\omega_{\vec{k}})^{1/2}}\bar{U}_{i}(\vec{p},s) T^{a}_{cc'}\gamma_{\mu}{U}_{j}(\vec{q},r)}\frac{1}{(-1)^{i+1}E_{\vec{p}}-(-1)^{j+1}E_{\vec{q}}+\omega_{\vec{k}}} \times \\
& \times{F^{c}_{i}}^{\dagger}(\vec{p},s) {F^{c'}_{j}}(\vec{q},r) {\epsilon^{\mu}}^{*}(\vec{k},\lambda)a^{\dagger}_{a}(\vec{k},\lambda) \delta(\vec{p}-\vec{q}+\vec{k}) - h.c.
\end{split}
\end{equation}

$R^{(1)}$ repeats the structure of interaction operator (\ref{field_hamiltonian11}), except it obtains energetic propagator. The structure of the propagator is itself reflects the structure of 3-linear interaction operator and presents energy imbalance in the interaction vertex. Since interaction $V$ determining $R^{(1)}$, has empty energy shell, energetic propagator in (\ref{field_hamiltonian3}) is always finite.

Now, it is possible to determine the explicit form of $[R^{(1)},V]$ and extract 'bad' terms of the 2nd order.
To obtain $3q\to3q$ potential in the 4th order, we need to consider $[R^{(1)},V]$ in the $FF\to FF$ sector, where $F$ is either quark of antiquark. It has the following form:
\begin{equation}\label{field_hamiltonian4}
\begin{split}
\frac{1}{2}[R^{(1)},V]_{FF\to FF} &= - \frac{1}{2}\frac{g^{2}}{(2\pi)^{3}}\int{\frac{d^{3}p_1d^{3}q_1d^{3}k_1d^{3}p_2d^{3}q_2d^{3}k_2}{(E_{\vec{p_1}}E_{\vec{q_1}}\omega_{\vec{k_1}})^{1/2}(E_{\vec{p_2}}E_{\vec{q_2}}\omega_{\vec{k_2}})^{1/2}}}
\delta(\vec{p_1}-\vec{q_1}+\vec{k_1})\delta(\vec{p_2}-\vec{q_2}+\vec{k_2})\times
\\ & \bar{U}_{i_1}(\vec{p_1},s_1) T^{a_1}_{c_1c'_1}\gamma_{\mu_1}{U}_{j_1}(\vec{q_1},r_1) \bar{U}_{i_2}(\vec{p_2},s_2) T^{a_2}_{c_2c'_2}\gamma_{\mu_2}{U}_{j_2}(\vec{q_2},r_2) \times
\\ & {\epsilon^{\mu_1}}(-\vec{k_1},\lambda_1){\epsilon^{\mu_2}}^{*}(\vec{k_2},\lambda_2) \frac{1}{(-1)^{i_2+1}E_{\vec{p_2}}-(-1)^{j_2+1}E_{\vec{q_2}}+\omega_{\vec{k}}} \times 
\\ & {F^{c_1}_{i_1}}^{\dagger}(\vec{p_1},s_1) {F^{c'_1}_{j_1}}(\vec{q_1},r_1){F^{c_2}_{i_2}}^{\dagger}(\vec{p_2},s_2) {F^{c'_2}_{j_2}}(\vec{q_2},r_2)\delta_{a_1a_2}\delta_{\lambda_1\lambda_2}\delta(\vec{k_1}+\vec{k_2})  + h.c.,
\end{split}
\end{equation}
'Good' part (Fig.\ref{good0}, left diagram) of (\ref{field_hamiltonian4}) corresponds to process $q\bar{q}\to q\bar{q}$. Along with that, operator (\ref{field_hamiltonian4}) contains a set of 'bad' terms (e.g., Fig. \ref{bad1}, both top and bottom left diagrams). They are either have empty energy shell or describe transitions between 'colored' states \textit{or} both. Let us consider the second type of them -- 'bad' operators in $[R^{(1)},V]$ with non-empty energy shell -- the part within $qq\to qq$ sector:
\begin{equation}\label{field_hamiltonian5}
\begin{split}
{\frac{1}{2}[R^{(1)},V]_{qq\to qq}} &= \frac{1}{2}\frac{g^{2}}{(2\pi)^{3}}\int{\frac{d^{3}p_1d^{3}q_1d^{3}p_2d^{3}q_2d^{3}k}{(E_{\vec{p_1}}E_{\vec{q_1}})^{1/2}(E_{\vec{p_2}}E_{\vec{q_2}})^{1/2}\omega_{\vec{k}}}}
\delta(\vec{p_1}-\vec{q_1}-\vec{k})\delta(\vec{p_1}-\vec{q_1}+\vec{k})\times
\\ & \bar{U}_{i_1}(\vec{p_1},s_1) T^{a_1}_{c_1c'_1}\gamma_{\mu_1}{U}_{j_1}(\vec{q_1},r_1) \bar{U}_{i_2}(\vec{p_2},s_2) T^{a_2}_{c_2c'_2}\gamma_{\mu_2}{U}_{j_2}(\vec{q_2},r_2) \times
\\ & {\epsilon^{\mu_1}}(\vec{k},\lambda){\epsilon^{\mu_2}}^{*}(\vec{k},\lambda) \frac{1}{E_{\vec{p_2}}-E_{\vec{q_2}}+\omega_{\vec{k}}} \times 
\\ & {b^{c_1}}^{\dagger}(\vec{p_1},s_1){b^{c_2}}^{\dagger}(\vec{p_2},s_2) {b^{c'_1}}(\vec{q_1},r_1) {b^{c'_2}}(\vec{q_2},r_2)\delta_{a_1a_2}  + h.c.,
\end{split}
\end{equation}

The part of generator $R^{2}$, responsible for eleminating 'bad' term (\ref{field_hamiltonian5}) from Hamiltonian, is found from (\ref{series1}) and has the following form:
\begin{equation}\label{field_hamiltonian6}
\begin{split}
R^{(2)}_{qq\to qq} &= \frac{1}{2}\frac{g^{2}}{(2\pi)^{3}}\int{\frac{d^{3}p_1d^{3}q_1d^{3}p_2d^{3}q_2d^{3}k}{(E_{\vec{p_1}}E_{\vec{q_1}})^{1/2}(E_{\vec{p_2}}E_{\vec{q_2}})^{1/2}\omega_{\vec{k}}}}
\delta(\vec{p_1}-\vec{q_1}-\vec{k})\delta(\vec{p_1}-\vec{q_1}+\vec{k})\times
\\ & \bar{U}_{i_1}(\vec{p_1},s_1) T^{a_1}_{c_1c'_1}\gamma_{\mu_1}{U}_{j_1}(\vec{q_1},r_1) \bar{U}_{i_2}(\vec{p_2},s_2) T^{a_2}_{c_2c'_2}\gamma_{\mu_2}{U}_{j_2}(\vec{q_2},r_2) \times
\\ & {\epsilon^{\mu_1}}(\vec{k},\lambda){\epsilon^{\mu_2}}^{*}(\vec{k},\lambda) \frac{1}{E_{\vec{p_2}}-E_{\vec{q_2}}+\omega_{\vec{k}}} \cdot \frac{1}{E_{\vec{p_2}}-E_{\vec{q_2}}+E_{\vec{p_1}}-E_{\vec{q_1}}} \times 
\\ & {b^{c_1}}^{\dagger}(\vec{p_1},s_1){b^{c_2}}^{\dagger}(\vec{p_2},s_2) {b^{c'_1}}(\vec{q_1},r_1) {b^{c'_2}}(\vec{q_2},r_2)\delta_{a_1a_2}  - h.c.,
\end{split}
\end{equation}

In this case, since process $qq\to qq$ has non-empty energy shell, the propagator of $R^{(2)}$ is not always finite. It means, that when we try to put operator (\ref{field_hamiltonian6}) on the energy shell, requiring $E_{\vec{p_2}}+E_{\vec{p_1}} = E_{\vec{q_2}}+E_{\vec{q_1}}$, it goes to infinity, indicating the color confinement of quarks.

Let us now derive the 'good' interaction potential, responsible for process $3q\to3q$, from the term $[R^{(2)},[R^{(1)},V]_{bad}]$.
For brevity, we make the following denotations:
\begin{equation}\label{notations6}
\begin{split}
\frac{1}{D_{1}} & = \frac{1}{E_{\vec{p_1}}-E_{\vec{q_1}}+\omega_{\vec{k_1}}}, \hskip 10mm \frac{1}{D^{*}_{1}} = \frac{1}{-E_{\vec{p_1}}+E_{\vec{q_1}}+\omega_{\vec{k_1}}}, \\
\frac{1}{\hat{D}_{1}} & = \frac{1}{E_{\vec{p_1}}+E_{\vec{q_1}}+\omega_{\vec{k_1}}}, \hskip 10mm \frac{1}{\hat{D}^{*}_{1}} = \frac{1}{-E_{\vec{p_1}}-E_{\vec{q_1}}+\omega_{\vec{k_1}}}, \\
\frac{1}{D_{12}} & = \frac{1}{E_{\vec{p_1}}-E_{\vec{q_1}}+E_{\vec{p_2}}-E_{\vec{q_2}}}, \frac{1}{D_{\hat{1}2}} = \frac{1}{E_{\vec{p_1}}+E_{\vec{q_1}}+E_{\vec{p_2}}-E_{\vec{q_2}}}\\
\frac{1}{D_{1^*2}} & = \frac{1}{-E_{\vec{p_1}}+E_{\vec{q_1}}+E_{\vec{p_2}}-E_{\vec{q_2}}}, \frac{1}{D_{\hat{1}^*2}} = \frac{1}{-E_{\vec{p_1}}-E_{\vec{q_1}}+E_{\vec{p_2}}-E_{\vec{q_2}}} \\
\end{split}
\end{equation}
\begin{equation}\label{notations7}
\begin{split}
K^{a_1}_{\mu_1, c_1c'_1}&(\vec{p_1}, i_1, \vec{q_1}, j_1) = \bar{U}_{i_1}(\vec{p_1},s_1) T^{a_1}_{c_1c'_1}\gamma_{\mu_1}{U}_{j_1}(\vec{q_1},r_1)
\end{split}
\end{equation}
In these notations, the term $\frac{1}{4}[R^{(2)}_{qq\to qq}, {[R^{(1)},V]_{qq\to qq}}]$ looks as follows:
\begin{equation}\label{field_hamiltonian7}
\begin{split}
\frac{1}{4}[R^{(2)}_{qq\to qq}, {[R^{(1)},V]_{qq\to qq}}] & = \frac{1}{8}\frac{g^4}{(2\pi)^6}\int{\frac{d^{3}p_1d^{3}k_2d^{3}k_4}{ E_{\vec{p_1}}\omega_{\vec{k_2}}\omega_{\vec{k_4}}  } } \cdot \frac{d^{3}q_1d^{3}p_2d^{3}q_2d^{3}p_3d^{3}q_3d^{3}p_4}{ (E_{\vec{q_1}}E_{\vec{p_2}}E_{\vec{q_2}}E_{\vec{p_3}}E_{\vec{q_3}}E_{\vec{p_4}})^{1/2} } \times
\\  K^{a_2}_{\mu_1, c_1c'_1}(\vec{p_1}, 1, \vec{q_1}, 1)& K^{a_2}_{\mu_2, c_2c'_2}(\vec{p_2}, 1, \vec{q_2}, 1)K^{a_4}_{\mu_3, c_3c'_3}(\vec{p_3}, 1, \vec{q_3}, 1)K^{a_4}_{\mu_4, c_4c_1}(\vec{p_4}, 1, \vec{p_1}, 1) \times
\\  {\epsilon^{\mu_1}}^{*}(\vec{k_2},\lambda_2)& {\epsilon^{\mu_2}}(\vec{k_2},\lambda_2) {\epsilon^{\mu_3}}^{*}(\vec{k_4},\lambda_4) {\epsilon^{\mu_4}}(\vec{k_4},\lambda_4) \times
\\  \delta(\vec{p_1}-\vec{q_1}-\vec{k_2})&\delta(\vec{p_2}-\vec{q_2}+\vec{k_2})\delta(\vec{p_3}-\vec{q_3}-\vec{k_4})\delta(\vec{p_4}-\vec{p_1}+\vec{k_4}) \times
\\ [\frac{2}{D_{34}}(\frac{1}{D_3}+\frac{1}{D^{*}_3})&(\frac{1}{D_1}+\frac{1}{D^{*}_1}+\frac{1}{D_2}+\frac{1}{D^{*}_2}) - \frac{2}{D_{12}}(\frac{1}{D_1}+\frac{1}{D^{*}_1})(\frac{1}{D_3}+\frac{1}{D^{*}_3}+\frac{1}{D_4}+\frac{1}{D^{*}_4}) \\
(\frac{1}{D_3D^*_4}-\frac{1}{D^{*}_3D_4})&(\frac{1}{D_1}+\frac{1}{D^{*}_1}+\frac{1}{D_2}+\frac{1}{D^{*}_2}) + (\frac{1}{D_1D^*_2}-\frac{1}{D^{*}_1D_2})(\frac{1}{D_3}+\frac{1}{D^{*}_3}+\frac{1}{D_4}+\frac{1}{D^{*}_4}) ]\times
\\ & {b^{c_4}}^{\dagger}(\vec{p_4},s_4){b^{c_3}}^{\dagger}(\vec{p_3},s_3){b^{c_2}}^{\dagger}(\vec{p_2},s_2){b^{c'_3}}(\vec{q_3},r_3){b^{c'_2}}(\vec{q_2},r_2){b^{c'_1}}(\vec{q_1},r_1),
\end{split}
\end{equation}

Adding to this operator the contribution from $3q\to3q$ operator with anti-quark in the intermediate state, we gather the covariant propagators, which include the following structures:
\begin{equation}\label{field_hamiltonian7}
\begin{split}
G_1^{3q\to3q}\sim\frac{1}{(p_4-q_3)(p_3-q_3)(p_3-q_3)^2(p_2-q_2)^2}, \\
G_2^{3q\to3q}\sim\frac{1}{(p_2-q_1)(p_2-q_2)(p_2-q_2)^2(p_3-q_3)^2}.
\end{split}
\end{equation}

The denotations for particle momenta are taken as in Figure \ref{good1}.
These potential is proportional to transferred momenta $(p_2-q_2)$ (or $(p_3-q_3)$) to the degree '-3'.
Compared to potentials that are proportional to the transferred momentum to '-2'nd degree.

This example is, of cause, only one of the family of confinement potentials, one of the easiest to explicitely calculate.
After defining 'color-clothed' representation of QCD Hamiltonian in all orders, we expect it (and thus the S-operator) -- though formally remaining the same -- to consist only of oparetors that describe transitions between 'color-neutral' states.

In addition, we also expect the 'clothing' procedure to affect mass and coupling constant renarmalization schemes. In particular, it is possible to expect that confinement potentials participate in corrections to quark masses and coupling constant in higher orders, effectively weakening coupling and driving quark masses somewhere close to constituent mass values.

Further investigation of confinement potentials properties, as well as renormalization scheme in concidered approach are subject to future studies.


\section{\label{sec:conclusions}Conclusions and plans}
Introducing the concept of 'color clothed' quarks and gluons, we suggested the approach to the effective QCD Hamiltonian, free of devergences and reflecting color confinement properties of quarks and gluons in an natural way, and showed simple example for calculation of quark-gluon potentials in this scheme.
Future plans are connected with deeper studies of confinement potentials' properties and also with renormalization scheme in 'color clothed' representation.
Any suggestions, comments or criticism would be appreciated very much.


\newpage
\begin{thebibliography}{99}
\bibitem{greenberg_1}
	\textit{Greenberg O. and Schweber S.} 
	Clothed particle operators in simple models of quantum field theory.
	\textit{Nuovo. Cim.} {1958, 8.}: pp. 378–405;
\bibitem{brodsky_1}
	\textit{Brodsky S.J., Pauli H.-C., and Pinsky S.S.} 
	Quantum Chromodynamics and Other Field Theories on the Light Cone. 
	\textit{Phys. Rep.} {1998, vol. 301, issues 4-6.}: pp. 299–486, arXiv:hep-ph/9705477;
\bibitem{shebeko_0}
	\textit{Shebeko A.V., Shirokov M.I.} 
	Unitary transformations in quantum field theory and bound states. 
	\textit{Phys. Part. Nucl.} {2001, vol. 32}: pp. 15–48, arXiv:nucl-th/0102037;
\bibitem{shebeko_1}
	\textit{Korda V.Yu., Canton L., and Shebeko A.V.} 
	Relativistic interactions for the meson-two-nucleon system in the clothed-particle unitary representation. 
	\textit{Ann. Phys.} {2007, vol. 322, issue 3.}: pp. 736–768, arXiv:nucl-th/0603025;
\bibitem{shebeko_2}
	\textit{Shebeko A.V.} 
	The Method of Unitary Clothing Transformations in Relativistic Quantum Field Theory: Recent Applications for the Description of Nucleon–Nucleon Scattering and Deuteron Properties. 
	\textit{Few-Body Syst.} {2013, vol. 54, issue 12.}: pp. 2271–2282;
\bibitem{stefanovich_1}
	\textit{Stefanovich E.V.} 
	Relativistic Quantum Dynamics: A non-traditional perspective on space, time, particles, fields, and action-at-a-distance. 
	\textit{arXiv:physics/0504062};
\bibitem{weinberg_1}
	\textit{Weinberg S.} 
	The quantum theory of fields. Vol.1. 
	\textit{Cambridge: University Press, 1995}, p. 350;
\end{thebibliography}
\end{document}